# Pressure-induced superconductivity in quasi-one-dimensional semimetal $Ta_2PdSe_6$

Haiyang Yang,[1,2] Yonghui Zhou,[1,]* Liangyu Li,[1,2] Zheng Chen,[1,2] Zhuyi Zhang,[1,2] Shuyang Wang,[1,2] Jing Wang,[1,2] Xuliang Chen,[1] Chao An,[3] Ying Zhou,[3] Min Zhang,[3] Ranran Zhang,[1] Xiangde Zhu,[1] Lili Zhang,[4] Xiaoping Yang,[1,]* and Zhaorong Yang[1,3,]*

[1] *Anhui Key Laboratory of Condensed Matter Physics at Extreme Conditions, High Magnetic Field Laboratory, HFIPS, Chinese Academy of Sciences, Hefei 230031, China*

[2] *Science Island Branch of Graduate School, University of Science and Technology of China, Hefei 230026, China*

[3] *Institutes of Physical Science and Information Technology, Anhui University, Hefei 230601, China*

[4] *Shanghai Synchrotron Radiation Facility, Shanghai Advanced Research Institute, Chinese Academy of Sciences, Shanghai 201204, China*

*Corresponding authors.
yhzhou@hmfl.ac.cn;
xpyang@hmfl.ac.cn;
zryang@issp.ac.cn

**Abstract**

Here we report the discovery of pressure-induced superconductivity in quasi-one-dimensional $Ta_2PdSe_6$, through a combination of electrical transport, synchrotron x-ray diffraction (XRD), and theoretical calculations. Our transport measurements show that the superconductivity appears at a critical pressure $P_c \sim 18.3$ GPa and is robust upon further compression up to 62.6 GPa. The estimated upper critical field $\mu_0 H_{c2}(0)$ in the pressurized $Ta_2PdSe_6$ is much lower than Pauli limiting field, in contrast to the case in its isostructural analogs $M_2Pd_xX_5$ (M = Nb, Ta; X = S, Se). Concomitant with the occurrence of superconductivity, anomalies in pressure dependent transport properties are observed, including sign reversal of Hall coefficient, abnormally enhanced resistance, and dramatically suppressed magnetoresistance. Meanwhile, room-temperature synchrotron X-ray diffraction experiments reveal the stability of the pristine monoclinic structure (space group $C2/m$) upon compression. Combined with the density functional theory calculations, we argue that a pressure-induced Lifshitz transition could be the electronic origin of the emergent superconductivity in $Ta_2PdSe_6$.

# I. INTRODUCTION

Low-dimensional transition metal chalcogenides (TMC) are an emerging class of materials with properties that make them highly attractive for fundamental studies of novel physical phenomena, and for potential applications in electronics and optoelectronics [1-5]. Among them, quasi-one-dimensional (quasi-1D) $M_2Pd_xX_5$ (M = Nb, Ta; X = S, Se) and $Ta_4Pd_3Te_{16}$ have been paid special attention, mainly focusing on their unusual superconductivity. On one hand, $M_2Pd_xX_5$ superconductors were reported to display an upper critical magnetic field $H_{c2}$ that is far beyond the Pauli paramagnetic limit [6-9]. The extremely large $H_{c2}$ could be ascribed to strong spin-orbit coupling, multi-band effect, or even spin-triplet pairing [6-11]. On the other hand, associated with their quasi-1D character, $M_2Pd_xX_5$ and $Ta_4Pd_3Te_{16}$ superconductors can be easily fabricated into nanowires, nanostrips and long fibers, leading to emergent phenomena such as size-controlled superconductor-insulator transition, and magnetic-field-induced reentrance of superconductivity [12-15]. Moreover, quasi-1D materials could have a strong tendency to form a charge density wave (CDW). For example, $Ta_4Pd_3Te_{16}$ was reported to display coexistence and interesting interplay between CDW and superconductivity [16-20].

As isostructural analogs of $M_2PdX_5$ (M = Nb, Ta; X = S, Se), $Ta_2PdX_6$ crystallize in monoclinic structure with the same space group $C2/m$ [21]. For example, $Ta_2PdSe_6$ features a layered structure oriented parallel to the plane $(20\bar{1})$, As shown in Fig. 1(a). The layer is composed of $PdSe_4$ quadrilaterals and $TaSe_7$ polyhedra. The only difference between $M_2PdX_5$ and $Ta_2PdX_6$ is that face-shared $TaX_7$ instead of $TaX_6$ prisms form the 1D chains along $b$-axis in $Ta_2PdX_6$. Since $Ta_2PdX_6$ have the very similar crystal structure as $M_2PdX_5$, it can be expected that $Ta_2PdX_6$ are also superconducting. Unfortunately, so far there is no definite experimental evidence to support for the superconductivity in $Ta_2PdX_6$ [22,23]. Instead, $Ta_2PdX_6$ have been studied as photoelectric and thermoelectric materials. For example, few-layer $Ta_2PdS_6$ was reported to exhibit excellent optoelectronic performances such as ultrahigh photoresponsivity and high photoconductivity [24]. Bulk $Ta_2PdSe_6$ was reported to host a giant Peltier conductivity and a large power factor at 300 K [25,26], and theoretically predicted to be a topological nodal-line semimetal candidate [27].

It is well known that pressure as an effective control parameter can systematically tune lattice structures and the corresponding electronic states. In this work, we show that $Ta_2PdSe_6$ can be driven into a superconductor under high pressure by electrical transport and synchrotron X-ray diffraction (XRD) measurements. The emergence of superconductivity at a critical pressure $P_c \sim 18.3$ GPa does not involve a structural transition, but is accompanied by abnormal evolution of normal state transport properties. Combining density functional theory (DFT) calculations and transport measurements, we conclude that a pressure-induced Lifshitz transition could be the electronic origin of the emergent superconductivity in $Ta_2PdSe_6$. Moreover, the $H_{c2}$ in the superconducting $Ta_2PdSe_6$ is much lower than the Pauli limiting field, which is different from that in $M_2PdX_5$ (M = Nb, Ta; X = S, Se).

## II. EXPERIMENTAL DETAILS & METHODS

$Ta_2PdSe_6$ single crystals were grown via chemical vapor transport method [25]. Stoichiometric amounts of high purity Ta, Pd, and Se with a weight of 0.5 g and 2 mg/cm$^3$ of iodide were mixed thoroughly and then sealed in a quartz tube with a low vacuum pressure of $\leq 10^{-3}$ Pa. The tube was heated in a horizontal tube furnace with a temperature gradient of 120°C between 800°C and 680°C for one week. The obtained high-quality single crystals are shiny with the needle shape, as seen in the inset of Fig. 1(b). Room-temperature XRD patterns of single crystal were obtained with Cu $K_\alpha$ radiation ($\lambda = 1.5406$ Å) using a Rigaku X-ray diffractometer (Miniflex-600). It indicates that the $(20\bar{1})$ plane is a natural cleavage facet. The atomic proportion of crystal was $Ta_{2.1}PdSe_{5.9}$, extracting from energy dispersive X-ray spectroscopy (EDXS, Helios Nanolab 600i, FEI) with area-scanning mode. The ambient pressure electrical transport measurements were performed in a physical property measurement system (PPMS, Quantum Design). Transport measurements were conducted along the *b*-axis of single crystal with a standard four-probe configuration, as shown in Fig. 1(d). The *R-T* curve is rather smooth and has no sharp anomaly, suggesting the absence of CDW transition in $Ta_2PdSe_6$. The sample exhibits a metallic behavior with a residual resistance ratio (RRR = $R_{300K}/R_{2K}$) of 1136.

High-pressure transport experiments were conducted in a screw-pressure-type diamond anvil cell made of CuBe alloy with a pair of anvil culets of 200 $\mu$m. The low-temperature electrical transport measurements were carried out in a homemade multifunctional measurement system (1.8-300 K, JANIS Research Company Inc.; 0-9 T, Cryomagnetics Inc.), while an additional He3 insert (with temperatures down to 0.45

K) provided the ultra-low temperature environment. In run 1 and run 2, five-probe resistance and Hall effect measurements were performed by sweeping the magnetic field from −5 to +5 T perpendicular to $(20\bar{1})$ plane with a steady current along the *b* axis. A mixture of epoxy and fine cubic boron nitride powder was compressed firmly to insulate the electrodes from the T301 stainless steel gasket. Then, a hole of 130 $\mu$m in diameter was drilled in the center of the pit, further filled with soft NaCl fine powder as the pressure-transmitting medium. A single crystal with dimensions of ~ 70 × 20 × 10 $\mu$m$^3$ was loaded together with some ruby powder. Platinum foil with a thickness of 5 $\mu$m was used for the electrodes. To detect the conductivity under an ultra-low temperature environment, a new round of electrical transport experiments was conducted with a standard four-probe configuration in run 3. The magnetic field orientation was consistent with runs 1 and run 2. The samples used in runs 1, run 2 and run3 are different pieces of single crystals from the same batch. High-pressure angle-dispersive synchrotron XRD experiments were carried out at room temperature with Ta$_2$PdSe$_6$ powder crushed from single crystals, at the beamline BL15U1 of the Shanghai Synchrotron Radiation Facility (SSRF) ($\lambda$ = 0.6199 Å). The diamonds with culet size of 300 $\mu$m and T301 stainless steel gasket were used. Daphne 7373 was used as the transmitting medium. A Mar345 image plate was used to record 2D diffraction patterns. The Dioptas [28] program was used for image integration, and the XRD patterns were fitted using the RIETICA [29] program with the Le Bail method. The pressure values for all of the above experiments were determined by the ruby fluorescence method at room temperature [30].

Electronic band structure and the Hellmann-Feynman force under different pressure were calculated by using the Vienna ab initio Simulation Package (VASP) [31,32] within the framework of generalized gradient approximation (GGA) (Perdew-Burke-Ernzerhof exchange functional) [33] and the modified Becke-Johnson (MBJ) exchange potential [34]. Spin-orbit coupling (SOC) for all elements was treated by a second-variation method. The ion-electron interaction was modeled by the projector augmented wave (PAW) method [35,36] with a uniform energy cutoff of 520 eV. Spacing between *k* points was 0.02 Å$^{-1}$. The geometry structures were optimized by employing the conjugate gradient technique, and in the final geometry, no force on the atoms exceeded 0.001 eV/Å. The phonon vibrational spectrum is investigated by using the finite displacement method implemented in Phonopy package [37]. In Fig. S1, GGA+SOC and GGA+MBJ+SOC band structures at 0 GPa are plotted. Our GGA+SOC band structure in Fig. S1(a) is consistent with that in Ref. [26]. The introduction of the MBJ potential in Fig. S1(b) further corrects the band structures and yields band gaps with similar accuracy to the hybrid functional results.

## III. RESULTS AND DISCUSSION

Figure 2 shows selected resistance-temperature $R(T)$ curves in the pressure range from 1.0 to 62.6 GPa. As shown in Fig. 2(a), the $R(T)$ curve at 1.0 GPa shares a similar metallic behavior with the case of ambient pressure. Below 12.8 GPa, $Ta_2PdSe_6$ remains metallic, and the resistance change slightly with increasing pressure. At 15.3 GPa, the $R(T)$ curve is shifted upward evidently. Upon further compression to 18.3 GPa, a tiny resistance drop is marked by the arrow at ~ 1.9 K, a temperature just above 1.8 K, the lowest temperature of the experiment in run 1. Such a drop becomes more and more pronounced at higher pressures, and the zero resistance is finally achieved above 38.1 GPa [see Fig. 2(b)], signaling the appearance of superconductivity in $Ta_2PdSe_6$. Furthermore, the superconductivity is very robust up to 62.6 GPa, the highest pressure conducted in this study. The pressure evolution of the transport property and the emergent superconductivity are reproducible in run 2, as shown in Fig. S2. The arrows indicate the onset of the superconducting transition. A slight upturn of resistance preceding the superconducting transition could be ascribed to the pressure inhomogeneity or pressure induced disorder.

To understand the pressure evolution of transport properties in $Ta_2PdSe_6$, high-pressure magnetoresistance (MR = $[(\rho_{xx}(H)-\rho_{xx}(0)]/\rho_{xx}(0)\times100\%$) and Hall resistivity $\rho_{xy}$ measurements were further conducted. Figure 3 shows the pressure-evolution of MR and Hall resistivity curves $\rho_{xy}(H)$ at 10 K under various pressures. These curves are plotted by the symmetric and anti-symmetric analyses, respectively, as shown in supplementary Fig. S3. Notably, the MR is strikingly suppressed with increasing pressure and is less than 1% above 18.3 GPa, which is comparable to $Nb_2Pd_xSe_5$ superconductor (less than 2% at 10 K) [7], as shown in the inset of Fig. 3(a). While the superconductivity is detected at 18.3 GPa, the MR effect becomes negligible in $Ta_2PdSe_6$. Note that the Weyl semimetal $WTe_2$ with large MR at ambient pressure becomes a superconductor at ~ 5 GPa, accompanied by the dramatically suppressed MR [38,39], which is quite similar to the case of pressurized $Ta_2PdSe_6$. As shown in

Fig. 3(b), the $\rho_{xy}(H)$ curve exhibits a nonlinear behavior with positive slope at 1.0 GPa, indicating a hole-dominated multiband feature. The slope of $\rho_{xy}(H)$ decreases monotonically with increasing pressure. At 15.3 GPa, the $\rho_{xy}(H)$ curve becomes to be linear in the whole field region. Further increase of the pressure above the critical pressure $P_c \sim 18.3$ GPa lead to the slope of $\rho_{xy}(H)$ changing from positive to negative. The sign reversal of the Hall resistivity around the pressure when the superconductivity starting to be observed is also reproducible in run 2 [see Fig. S4], demonstrates a crossover of the dominant carrier type concomitant with the emergent superconductivity. We noted that the electron-dominated behavior observed in pressurized $Ta_2PdSe_6$ (above $P_c$) is similar to the case of $Nb_2PdS_5$ [40] and $Nb_2Pd_xSe_5$ [7] superconductors.

The effect of the external magnetic field on superconductivity of pressurized $Ta_2PdSe_6$ was investigated in run 3. As shown in Fig. 4(a), when the temperature cooled down to 0.5 K, the zero-resistance state was observed at 17.7 GPa, signaling the appearance of superconductivity in $Ta_2PdSe_6$. The $T_c^{zero}$ is enhanced monotonously with increasing pressure. As seen in Fig. 4(b), the superconducting transition is gradually suppressed with increasing magnetic field. To determine the upper critical field $\mu_0H_{c2}(0)$, we chose a resistance criterion of $R_{cri} = 90\%R_n$ ($R_n$ is the normal state resistance) and plotted the temperature dependence of $\mu_0H_{c2}$ in Fig. 4(d). The Werthamer–Helfand–Hohenberg (WHH) model was used to fit the data [41]. The estimated $\mu_0H_{c2}(0) \sim 2.1$ T is much lower than the weak-coupling Pauli limit $\mu_0H_P(0) = 1.84T_c$ ($\sim 3.8$ T), which suggests the absence of Pauli paramagnetic pair-breaking effect. According to the relationship $\mu_0H_{c2} = \Phi_0/(2\pi\xi^2)$, where $\Phi_0 = 2.07\times10^{-15}$ Wb is the flux quantum, the coherence length $\xi_{GL}(0)$ of 125.5 Å is obtained. Similar analyses were applied to the case of 39.7 GPa. The obtained $\mu_0H_{c2}(0)$ and $\xi_{GL}(0)$ are 2.7 T and 110.7 Å, respectively. In contrast with $Ta_2PdSe_6$, the ratio of $\mu_0H_{c2}(0)/T_c$ were about 5.6-6.2 in $M_2Pd_xX_5$ (M = Nb, Ta; X = S, Se) superconductors, much larger than that of Pauli limit $\sim 1.84$ [6-9]. The significant difference in the upper critical field value suggested that the superconducting mechanism is likely different in superconducting $Ta_2PdSe_6$ and $M_2Pd_xX_5$.

To clarify the relationship between the emergent superconductivity and crystal structure in pressurized $Ta_2PdSe_6$, we further performed high-pressure synchrotron XRD experiments. As shown in Fig. 5(a), all the XRD peaks continuously shift towards higher angles without appearance of new peaks up to 36.2 GPa. Using the Le Bail method, the XRD patterns at each pressure can be indexed with the monoclinic $C2/m$ structure at ambient pressure [21]. Note that the strongest peak is (111), rather than $(20\bar{1})$ from the calculated XRD pattern, as shown in Fig. S5. The different relative strength of the XRD peaks between the experiment and calculation could be ascribed to the preferred orientation in pressurized $Ta_2PdSe_6$ powder. Typical analysis of the XRD patterns at 0.4, 11.2, 21.8, and 30.5 GPa are shown in supplementary Fig. S6. We extract the lattice parameters as a function of pressure, as shown in Fig. 5(b) and supplementary Table 1. The pressure dependent volume is depicted in Fig. 5(c). By the fitting of the third-order Birch-Murnaghan equation of state [42], we obtained the ambient pressure volume $V_0$ = 391.1 ± 1.8 Å$^3$, bulk modulus $B_0$ = 50.5 ± 4.7 GPa, and its first pressure derivative $B_0$' = 8.4 ± 0.2, respectively. Moreover, we examined the phonon stability of the $C2/m$ structure under pressure. As shown in supplementary Fig. S7, $Ta_2PdSe_6$ is dynamically stable under pressure up to 40 GPa in the monoclinic structure. These results indicate no structure phase transition under pressure, thus ruling out the structural origin of the pressure-induced superconductivity in $Ta_2PdSe_6$.

To have a comprehensive understanding of the superconductivity in the pressurized $Ta_2PdSe_6$, we constructed a phase diagram in Fig. 6. There are two distinct regions separated by a critical pressure $P_c$ ~ 18.3 GPa, *i.e.*, the semimetallic state with MR and the superconducting state. It is clear that the appearance of superconductivity is accompanied by unusual evolution of the transport properties, including the dramatically suppressed MR, abrupt enhancement of resistance at 10 K, and sign change of the Hall coefficient $R_H$ [see Fig. 6(b)]. With increasing pressure, it is rare to observe concurrently the large change in the MR, Hall coefficient and resistance of a common compound. Without detecting a structural transition, the simultaneously occurrence of anomalies in the transport properties is reminiscent of a Lifshitz transition [43-50]. Such transition is a peculiar "electron transition" due to variation of the topology of the Fermi surface during its continuous deformation at high pressures [43]. Especially, the sign reversal of the $R_H$ indicates that the hole-dominated transport behavior maintains up to 18.3 GPa and transforms into electron-dominated behavior at higher pressures, which could be viewed as a signature of a modification of the Fermi surface. We note that the reported Lifshitz transitions in $ZrTe_5$ and $NiTe_2$ also involve the change of charge carrier type [44,45], which are very similar to our case. As shown in Fig. 6(a), upon compression, the $T_c^{zero}$ of the pressurized $Ta_2PdSe_6$ increases and the superconductivity is robust up to 62.6 GPa, in contrast to a dome-shaped phase diagram in quasi-1D $Ta_4Pd_3Te_{16}$ due to the competition between CDW and superconductivity [19,20].

Furthermore, we calculated pressure-dependent GGA+MBJ+SOC band structures at 0, 10, 20, 30, and 40 GPa, and the results are shown in Fig. 7. In the case of pressurized $Ta_2PdSe_6$, Fermi surface topology has relatively large changes along with the increment of pressure. Two significant evolutions are found in the electronic band structure. First, three conduction bands around $I|I_1$, $X_1|X$, and $N|M$ points are downshifted below the Fermi Level starting from 10 GPa, leading to a Lifshitz transition for the Fermi surface topology (Fig. 7(b-e)). Second, a valence band (marked by a black arrow) around Γ-point is lifted above the Fermi Level at 20 GPa (Fig. 7(c-e)), demonstrating a Lifshitz transition associated with the change of the Fermi surface topology around Γ-point. The Lifshitz transition could be a driven force for the appearance of superconductivity in pressurized $Ta_2PdSe_6$. Pressure-induced superconductivity has been associated with the Lifshitz transition in the systems such as the iron-based superconductors, black phosphorus, $PtTe_2$, and topological kagome metal $Cs_3VSb_5$ when there exists no structural phase transition [46-50]. For instance, it is observed that superconductivity reemerges at 5 GPa in FeS superconductor [46], which is theoretically ascribed to a Lifshitz transition [47].

## IV. CONCLUSIONS

In conclusion, we have systematically investigated the pressure effect on the structural and electronic properties of quasi-1D semimetal $Ta_2PdSe_6$. We observed a superconducting transition above the critical pressure $P_c \sim 18.3$ GPa. In contrast to isostructural $Ta_2PdSe_5$, we found the absence of Pauli paramagnetic pair-breaking effect in the pressurized $Ta_2PdSe_6$. Despite of the robust pristine monoclinic structure under pressure, abnormal evolutions of transport properties occur concurrently around $P_c$, implying an electronic origin of the emergent superconductivity. We hope our results will stimulate further theoretical work to address the nature of the superconductivity in $Ta_2PdSe_6$.

## ACKNOWLEDGMENTS

The authors gratefully acknowledge financial support from the National Key Research and Development Program of China (Grant Nos. 2018YFA0305700, 2021YFA1600204), the National Natural Science Foundation of China (Grant Nos. U1932152, 11874362, 12174395, 12004004, 11704387, U19A2093, U1832209, and 12174397), the Natural Science Foundation of Anhui Province (Grant Nos. 1808085MA06, 2008085QA40, and 1908085QA18), the Users with Excellence Project of Hefei Center CAS (Grant Nos. 2021HSC-UE008 and 2020HSC-UE015), the Collaborative Innovation Program of Hefei Science Center CAS (Grant No. 2020HSC-CIP014). A portion of this work was supported by the High Magnetic Field Laboratory of Anhui Province under Contract Nos. AHHM-FX-2020-02 and AHHM-FX-2021-03. Yonghui Zhou was supported by the Youth Innovation Promotion Association CAS (Grant No. 2020443). The high-pressure synchrotron X-ray diffraction experiments were performed at the beamline BL15U1, Shanghai Synchrotron Radiation Facility.

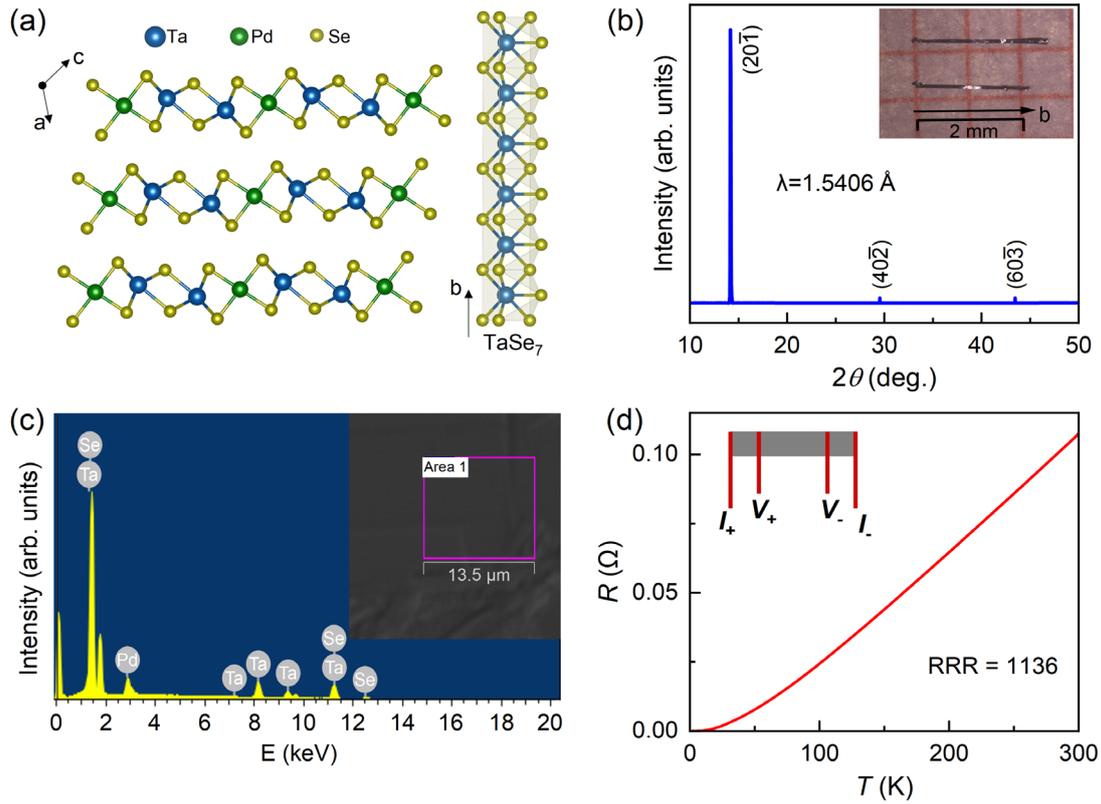

FIG. 1. (a) Schematic crystal structure of Ta$_2$PdSe$_6$ (monoclinic, space group $C2/m$). The blue, green, and orange spheres represent Ta, Pd, and Se, respectively. (b) XRD pattern of Ta$_2$PdSe$_6$ single crystal at room temperature ($\lambda$ = 1.5406 Å). The inset shows a picture of as-grown single crystals. (c) Energy-dispersive X-ray spectroscopy of a piece of single crystal with area-scanning mode. (d) Temperature dependence of resistance of a piece of single crystal with a residual resistance ratio RRR = 1136 at ambient pressure. Inset: Schematic configuration of the four-probe electrical transport measurement.

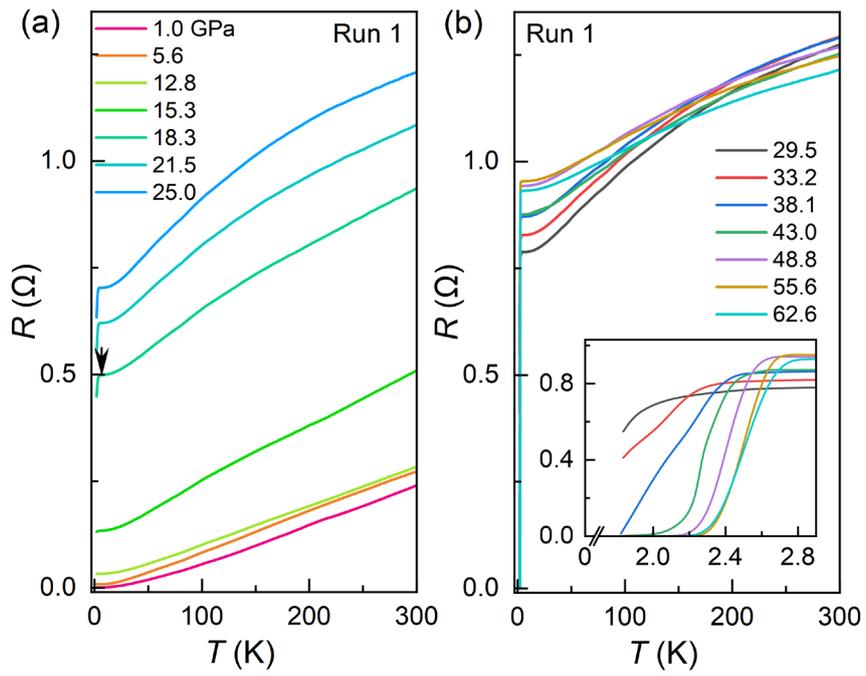

FIG. 2. Pressure evolution of resistance-temperature curves in Ta$_2$PdSe$_6$. Inset of (b): Enlarged view of the low-temperature region in run 1.

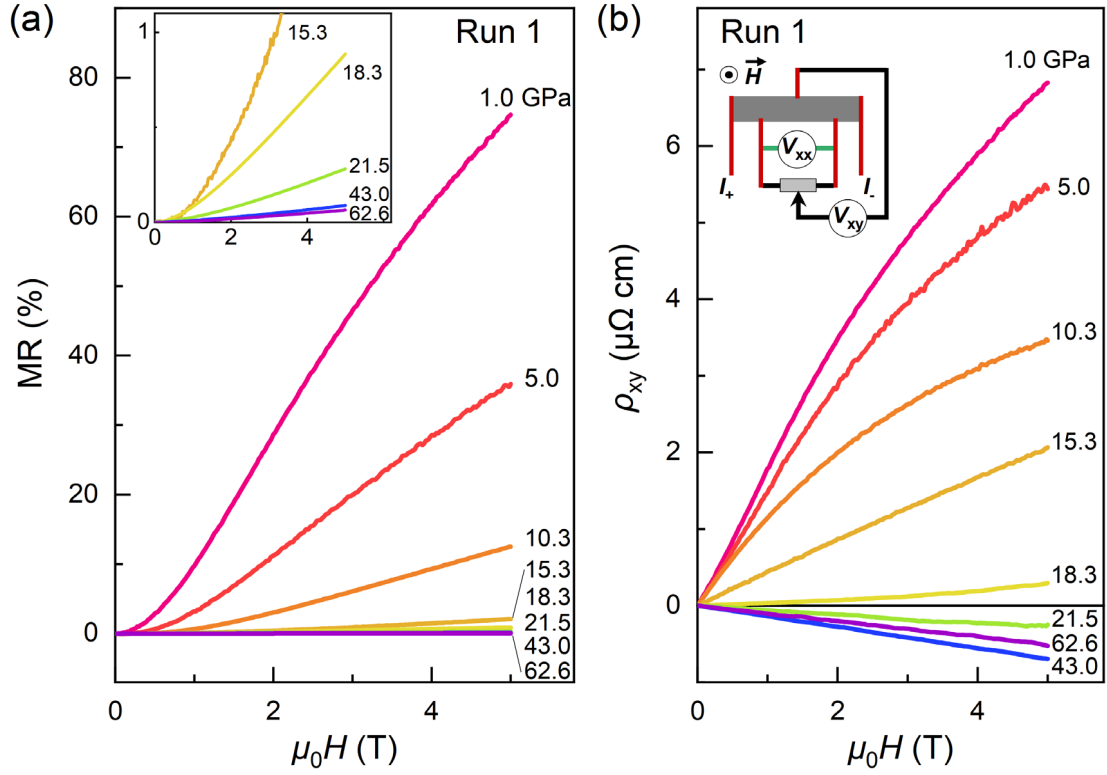

FIG. 3. (a) Pressure evolution of representative magnetoresistance curves in $Ta_2PdSe_6$ at 10 K. Inset: Enlarged view of the pressure above 15.3 GPa in run 1. (b) Pressure evolution of representative Hall resistivity curves $\rho_{xy}(H)$ in $Ta_2PdSe_6$ at 10 K. Inset: Schematic configuration of the electrical transport measurement with five probes.

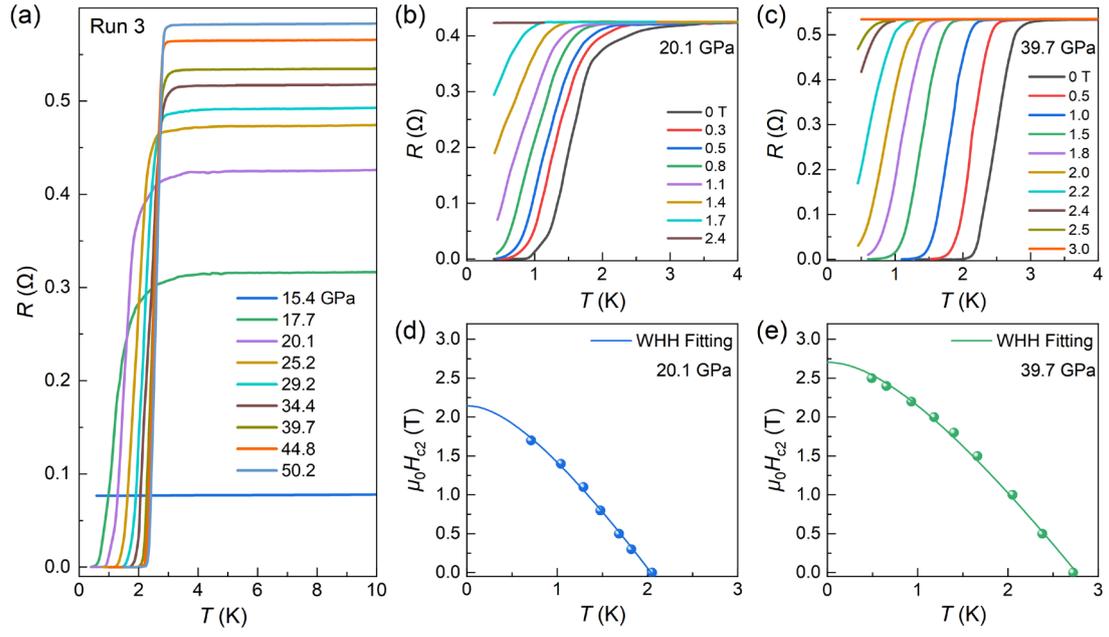

FIG. 4. Pressure evolution of resistance-temperature curves in run 3. (a) Enlarged view of the low-temperature region. Temperature-dependent resistance curves in Ta$_2$PdSe$_6$ under various magnetic fields at (b) 20.1 GPa and (d) 39.7 GPa in run 3. (c,e) Temperature dependence of the upper critical field $\mu_0H_{c2}$. The solid lines represent the WWH fitting.

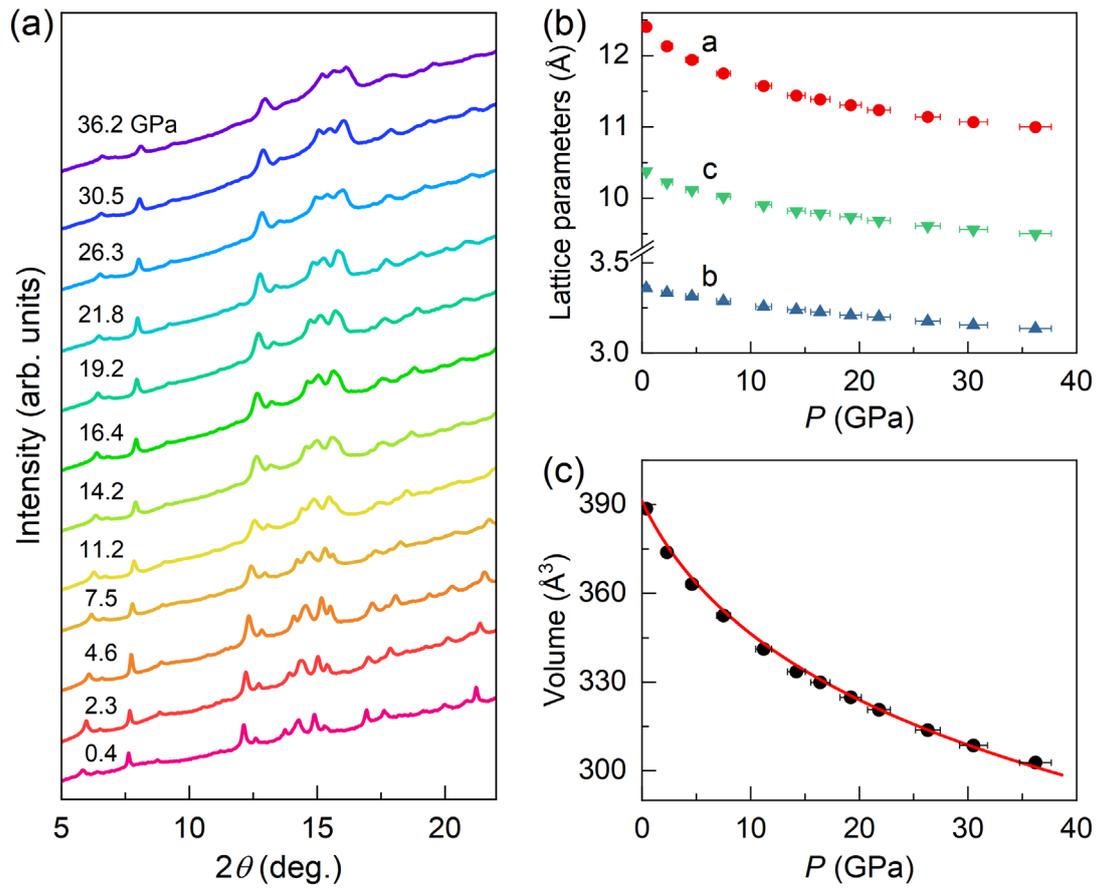

FIG. 5. (a) Pressure dependence of XRD patterns of $Ta_2PdSe_6$ at room temperature ($\lambda$ = 0.6199 Å). (b) Lattice parameters *a*, *b*, and *c* as a function of pressure. (c) Volume-pressure phase diagram. The solid red line is fits based on the third-order Birch-Murnaghan equation of states.

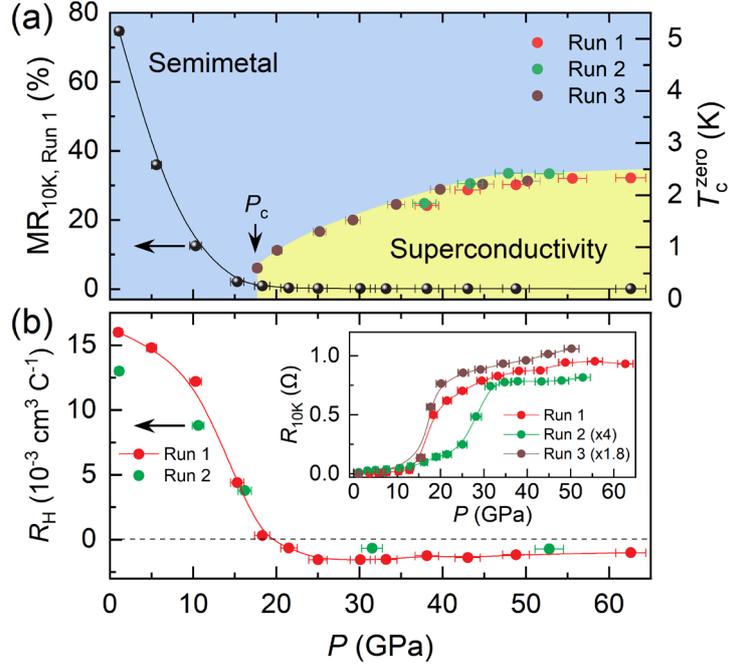

FIG. 6. Pressure–temperature phase diagram of $Ta_2PdSe_6$ and pressure-dependent Hall coefficient. (a) The red, green, and brown solid dots represent $T_c^{zero}$ for run 1, run 2, and run 3, respectively. (b) Hall coefficient ($R_H$) as a function of pressure measured at 10 K and 5 T. Inset: Pressure-dependent resistance of $Ta_2PdSe_6$ at 10 K for three runs.

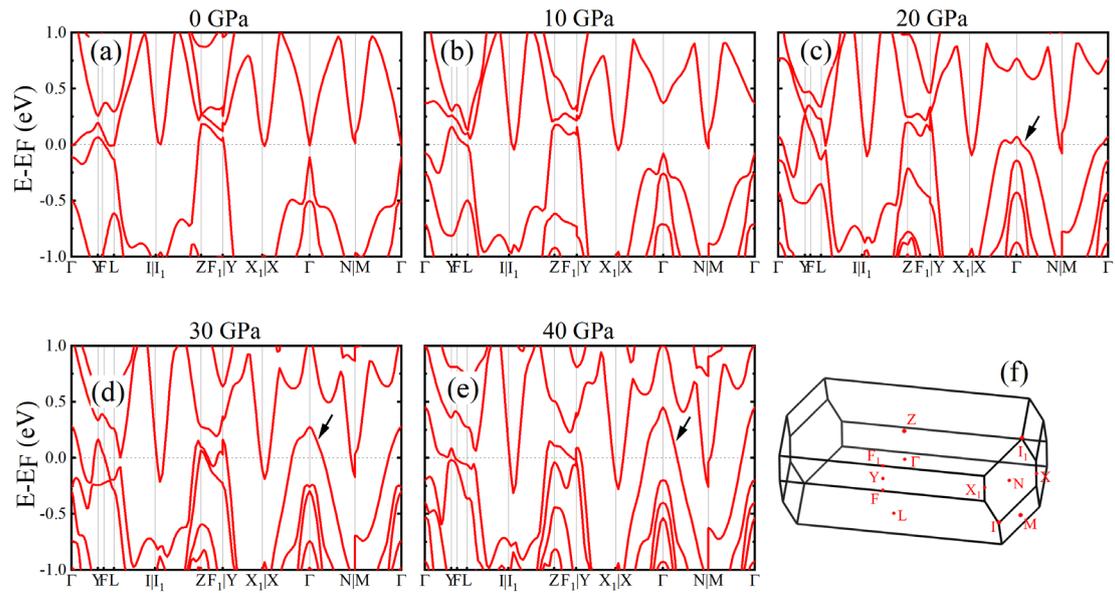

FIG. 7. (a)-(e) Calculated GGA+MBJ+SOC band structure at different pressure points (0, 10, 20, 30 and 40 GPa). Pressure dependent evolution of hole band pocket around Γ-point (marked by black arrow in (c)-(e)). (f) High symmetry points of the Brillouin zone.

Supplemental Material for

# Pressure-induced superconductivity in quasi-one-dimensional semimetal Ta$_2$PdSe$_6$

Haiyang Yang,[1,2] Yonghui Zhou,[1,*] Liangyu Li,[1,2] Zheng Chen,[1,2] Zhuyi Zhang,[1,2] Shuyang Wang,[1,2] Jing Wang,[1,2] Xuliang Chen,[1] Chao An,[3] Ying Zhou,[3] Min Zhang,[3] Ranran Zhang,[1] Xiangde Zhu,[1] Lili Zhang,[4] Xiaoping Yang,[1,*] and Zhaorong Yang[1,3,*]

[1] *Anhui Key Laboratory of Condensed Matter Physics at Extreme Conditions, High Magnetic Field Laboratory, HFIPS, Chinese Academy of Sciences, Hefei 230031, China*

[2] *Science Island Branch of Graduate School, University of Science and Technology of China, Hefei 230026, China*

[3] *Institutes of Physical Science and Information Technology, Anhui University, Hefei 230601, China*

[4] *Shanghai Synchrotron Radiation Facility, Shanghai Advanced Research Institute, Chinese Academy of Sciences, Shanghai 201204, China*

*Corresponding authors.
yhzhou@hmfl.ac.cn;
xpyang@hmfl.ac.cn;
zryang@issp.ac.cn

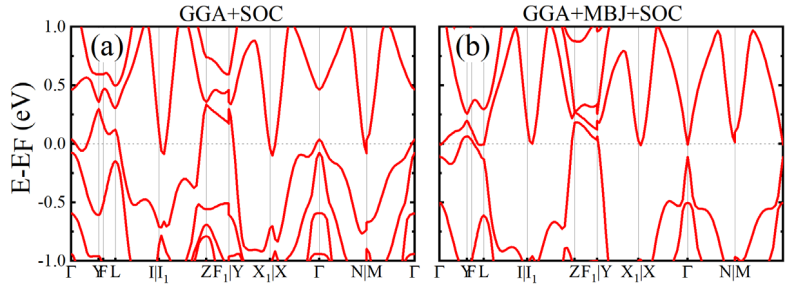

Fig. S1 (a,b) Calculated GGA+SOC and GGA+MBJ+SOC band structure at 0 GPa.

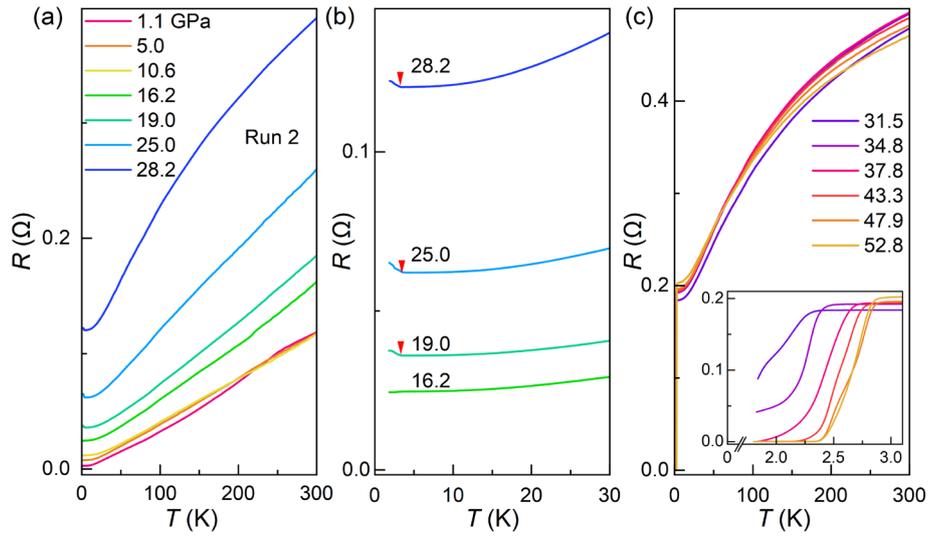

FIG. S2. Pressure evolution of resistance-temperature curves in Ta$_2$PdSe$_6$ in run 2. The arrows in (b) indicates the onset of the superconducting transition. Inset of (c): Enlarged view of the low-temperature region.

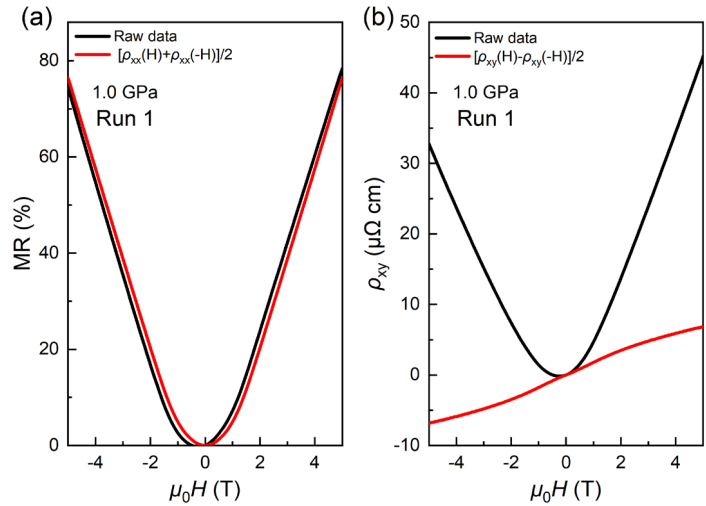

FIG. S3. Magnetoresistance (MR = [($\rho_{xx}(H)-\rho_{xx}(0)$)/$\rho_{xx}(0)\times100\%$) and Hall resistivity curves at 10 K and 1.0 GPa in run 1 are plotted by the symmetric and anti-symmetric analyses, respectively.

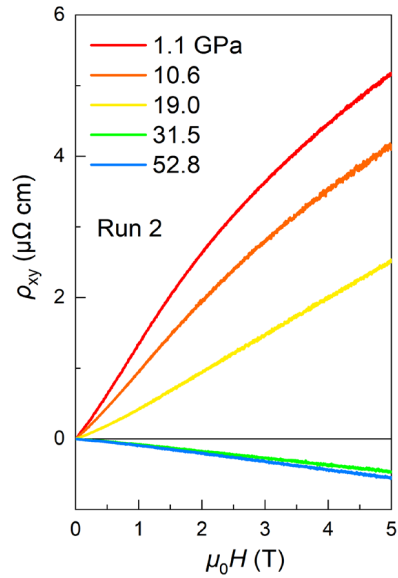

FIG. S4. Pressure evolution of representative Hall resistivity curves $\rho_{xy}(H)$ at 10 K in run 2.

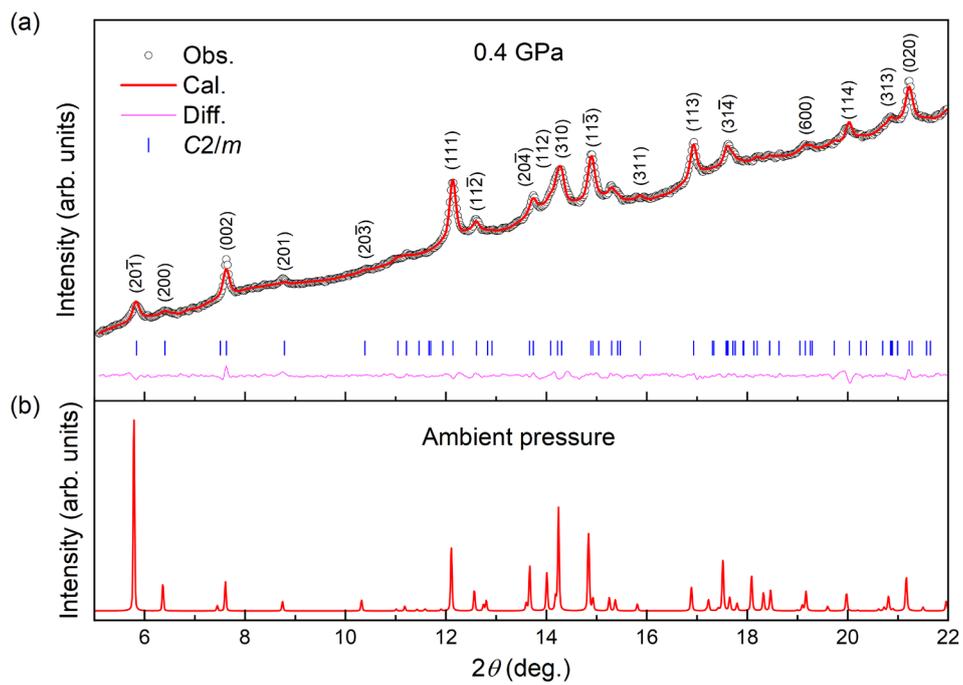

FIG. S5. (a) Representative Le Bail refinement of the XRD patterns at 0.4 GPa. The open circles and solid lines represent the observed and calculated data, respectively. The solid lines at the bottom denote the residual intensities. The vertical bars indicate the Bragg peak positions with monoclinic $C2/m$ phase. (b) The calculated XRD pattern at ambient pressure.

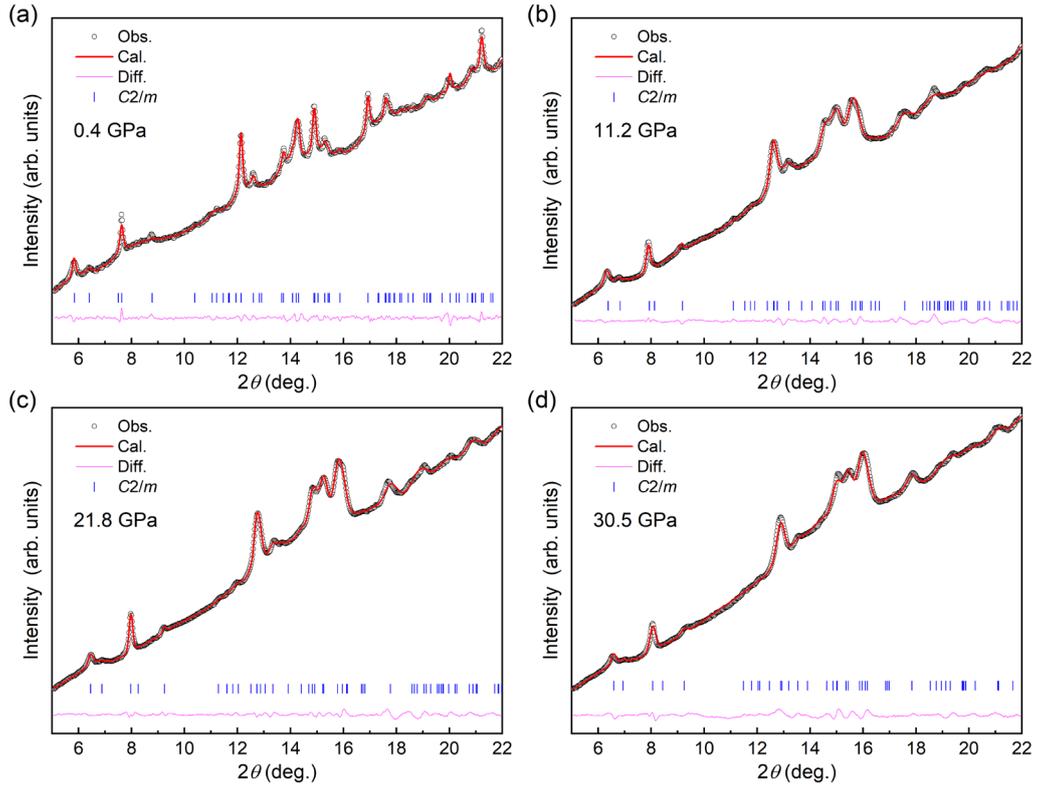

FIG. S6. Representative Le Bail refinements of the XRD patterns at 0.4, 11.2, 21.8, and 30.5 GPa, respectively. The open circles and solid lines represent the observed and calculated data, respectively. The solid lines at the bottom denote the residual intensities. The vertical bars indicate the Bragg peak positions with monoclinic $C2/m$ phase.

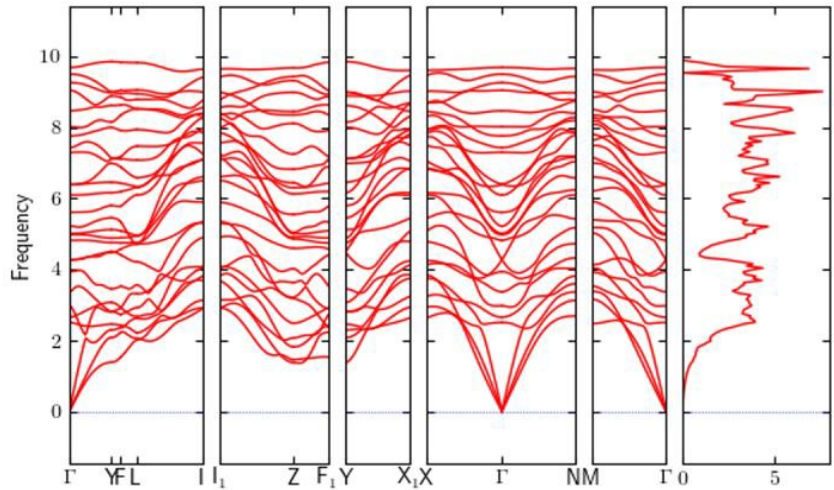

FIG. S7. The calculated phonon spectra of $Ta_2PdSe_6$ at 40 GPa.

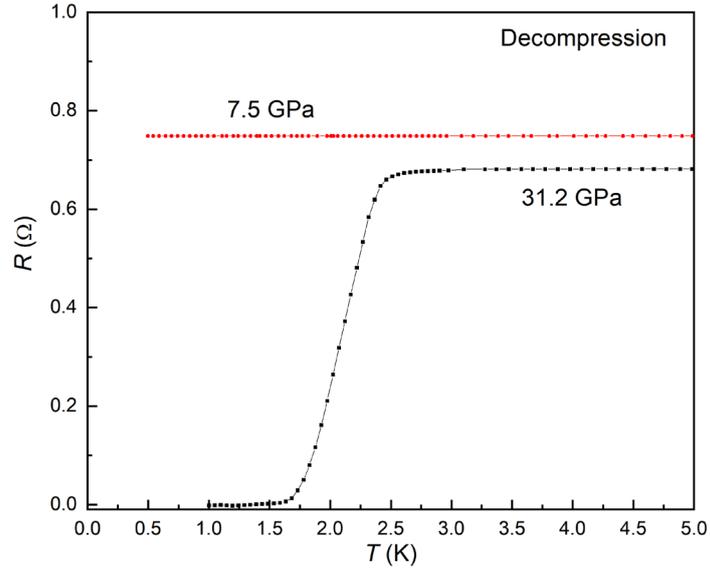

FIG. S8. $R(T)$ curve at 31.2 and 7.5 GPa during the decompression in run 3.

Table 1. The extracted lattice parameters as a function of pressure at room temperature.

| $P$(GPa) | $a$(Å) | $b$(Å) | $c$(Å) | $V$(Å$^3$) | $R_p$ | $R_{wp}$ |
|---|---|---|---|---|---|---|
| 0.4 | 12.402 | 3.360 | 10.377 | 388.624 | 0.7 % | 1.3 % |
| 2.3 | 12.131 | 3.334 | 10.224 | 373.916 | 0.6 % | 1.2 % |
| 4.6 | 11.939 | 3.314 | 10.117 | 363.159 | 0.7 % | 1.4 % |
| 7.5 | 11.748 | 3.288 | 10.018 | 352.51 | 0.6 % | 1.1 % |
| 11.2 | 11.573 | 3.256 | 9.903 | 341.196 | 0.6 % | 0.9 % |
| 14.2 | 11.437 | 3.240 | 9.817 | 333.546 | 0.6 % | 1.0 % |
| 16.4 | 11.384 | 3.226 | 9.785 | 329.908 | 0.6 % | 1.2 % |
| 19.2 | 11.303 | 3.210 | 9.737 | 324.836 | 0.5 % | 1.1 % |
| 21.8 | 11.236 | 3.199 | 9.686 | 320.649 | 0.7 % | 1.2 % |
| 26.3 | 11.138 | 3.176 | 9.611 | 313.784 | 0.5 % | 1.0 % |
| 30.5 | 11.067 | 3.156 | 9.558 | 308.551 | 0.6 % | 0.8 % |
| 36.2 | 10.996 | 3.134 | 9.498 | 302.780 | 0.6 % | 1.1 % |